%% file: main.tex
\documentclass[pra,floatfix,footinbib,onecolumn]{revtex4}

\input{header_preserve}
\usepackage{amsmath,amssymb,mathtools}
\usepackage[euler]{textgreek}
\usepackage[bbgreekl]{mathbbol}
\usepackage{graphicx}
\usepackage{bm,color}
\usepackage{array}
\usepackage{natbib,hyperref}
\usepackage[capitalize]{cleveref}

\newcommand{\fnpl}{{\scalebox{0.75}{$+$}}}
\newcommand{\fnpm}{{\scalebox{0.75}{$-$}}}

\hypersetup{breaklinks=true,colorlinks=true,linkcolor=blue,urlcolor=blue,citecolor=blue}

\input{header_command.tex}

\begin{document}
\title{Mapping repulsive to attractive interaction in 
driven-dissipative quantum systems}
\author{Andy C.~Y.~Li}
\affiliation{Department of Physics and Astronomy, Northwestern University, Evanston, Illinois 60208, USA}
\author{Jens Koch}
\affiliation{Department of Physics and Astronomy, Northwestern University, Evanston, Illinois 60208, USA}
\date{\today}

\begin{abstract}
Repulsive and attractive interactions usually lead to very different physics. 
Striking exceptions exist in the dynamics of driven-dissipative 
quantum systems. 
For the example of a photonic Bose-Hubbard dimer, we establish a one-to-one 
mapping relating the cases of onsite repulsion and attraction.
We prove that 
the mapping is valid for an entire class of Markovian open quantum systems with 
time-reversal invariant 
Hamiltonian and physically meaningful inverse-sign Hamiltonian.  To underline 
the broad 
applicability of the mapping, we illustrate the one-to-one correspondence 
between the 
nonequilibrium dynamics in a geometrically frustrated spin lattice and that in 
a 
non-frustrated partner lattice.
\end{abstract}

\maketitle

\section{Introduction}
Photonic quantum systems provide a versatile platform to study 
nonequilibrium many-body phenomena of light 
\cite{Grujic2013,Higgins2014,Raftery2014,Klinder2015,Hamel2015}, dissipative 
phase transitions \cite{Lee2012, 
	Kessler2012,Carmichael2015,Casteels2017,Fink2017,Fitzpatrick2017},  and 
	dissipation 
engineering 
\cite{Kraus2008,Sweke2013,Reiter2013,Kronwald2014,Aron2014,Kimchi-Schwartz2016}.
The nonequilibrium dynamics and steady-state properties of driven-dissipative 
systems also play a crucial role in the development of quantum information 
technology for quantum optimal control and open-system state stabilization 
\cite{Diehl2008,Schulte2011,Reiter2013,Kimchi-Schwartz2016}.
Despite the immense theoretical and experimental progress in this field, 
understanding the 
dynamics at an intuitive level often remains challenging. Considerations 
based on energetically favorable states are not generally appropriate in 
nonequilibrium, and 
can in fact be misleading.

We show that this is, in particular, the case for a driven-dissipative 
Bose-Hubbard dimer. 
Specifically, suppose that bosonic excitations are fed coherently into a dimer 
site, where 
they are subject to hopping, onsite interaction, and dissipation. How will the 
physics 
change when the sign of the onsite interaction is swapped, so that onsite 
repulsion turns into 
onsite attraction? In this paper, we demonstrate that there is an exact mapping 
relating 
observable expectation values for the repulsive system to those of the 
attractive system. In 
other words, while the equilibrium physics of a Bose-Hubbard dimer with 
conserved particle 
number is extremely different for attraction vs.\ repulsion \cite{Oelkers2007}, 
we find that the nonequilibrium 
dynamics of a driven-dissipative Bose-Hubbard dimer essentially does not 
distinguish between 
the repulsive and attractive case.

The mapping can be generalized and holds for a large class of Markovian open 
quantum 
systems with 
time-reversal invariant system Hamiltonian. It relates the 
nonequilibrium dynamics of 
an open system $Q_1$, associated with Hamiltonian $\oH$, to that of another 
system $Q_2$, 
associated with the negative-partner Hamiltonian $-\oH$. As long as $-\oH$ has 
physical 
meaning (e.g., as an effective Hamiltonian in a rotating frame), the 
mapping guarantees 
a one-to-one correspondence between observable expectation values 
in two different quantum systems $Q_1$ and $Q_2$.

In the remainder of this paper, we first discuss the Bose-Hubbard dimer model 
with drive and 
dissipation, approaching the mapping from the point of view of the equations 
of motion for 
system observables. We then prove that the result is an instance of the more 
general mapping, namely the Hamiltonian sign inversion (HSI) mapping,
which is applicable to a broad range of driven-dissipative systems. We 
illustrate this point 
by the discussion of another example, namely the mapping of the spin dynamics 
in a 
geometrically 
frustrated lattice to corresponding dynamics in a non-frustrated spin lattice.

\section{Driven-dissipative Bose-Hubbard dimer: mapping positive to negative $U$}
\label{sec:BH_dimer}
We consider a driven-dissipative 
Bose-Hubbard dimer \cite{Graefe2014,Cao2016,Casteels2017b,Casteels2017c} with 
either repulsive or 
attractive onsite interaction, $U>0$ or $U<0$, respectively. By inspection of 
the equations of motion, we will reveal an exact mapping between the cases of 
positive and 
negative $U$, i.e., between dimers with repulsive and attractive onsite 
interaction. The notion of such 
a mapping may, at first, seem to contradict the common intuition 
that attraction and repulsion must lead to entirely different physics. In our 
following derivation of the HSI mapping for the driven-dissipative Bose-Hubbard 
dimer, we 
will carefully discuss how this contradiction is resolved, and what exactly the 
mapping does and does not imply.

In concrete terms, the Bose-Hubbard dimer is described by the Hamiltonian
\begin{align}
\label{eq:H_dimer}
\oH_{\pm} = &
\sum_{n=1}^{2} \lf( \omr \, \oa_n\dg \oa_n  + U_\pm \oa_n\dg \oa_n\dg \oa_n \oa_n 
\rt) + J \lf( \oa_1\dg \oa_2 + \oa_2\dg \oa_1 \rt),
\end{align}
and consists of two sites, $n=1,\,2$, with onsite energy $\omr$ (we set 
$\hbar=1$ throughout) 
and Hubbard interaction of strength $U_\pm \gtrless 0$. Equation 
\eqref{eq:H_dimer} captures both the repulsive case (positive $U$) via 
$\oH_{+}$, and the attractive case (negative $U$) via $\oH_{-}$. Bosonic 
excitations are created  by $\oa_n\dg$ and can hop between the two sites 
$n=1,2$ 
with rate $J$. 

Simple energetic considerations suggest that repulsion and attraction lead to 
rather different results: 
In the positive-$U$ dimer, bosons repel: the onsite interaction $U_+ \oa_n\dg 
\oa_n\dg \oa_n \oa_n$ increases the energy quadratically with the number of bosons on each 
site. For fixed boson number $N\gg1$, the onsite interaction is minimized by 
 dividing the boson number equally between the two sites. By contrast, in the 
negative-$U$ dimer, bosons attract: the Hubbard term $U_- \oa_n\dg \oa_n\dg 
\oa_n \oa_n$ lowers the energy quadratically with the number of bosons on each 
site. For fixed boson number $N\gg1$, the onsite energy can thus be 
minimized by having all bosons occupy the same site. 
(We note that the spectrum of $\oH_-$ is not bounded from 
below if the boson number is not fixed: ultimately, adding more and more bosons will lower 
the energy indefinitely.  In practice, the attractive Bose-Hubbard dimer may serve as an 
\emph{effective} model, in which additional 
nonlinear interactions need to be included when the boson number exceeds a 
certain threshold. Such additional terms restoring boundedness of the spectrum 
will naturally be system dependent \footnote{For example when using the attractive 
Bose-Hubbard model as an approximation of a transmon qubit, excitations with energies above 
the maximum of the cosine potential will break the Bose-Hubbard approximation \cite{Koch2007}. 
In that case, a perturbative treatment of the potential is not appropriate.}.)

The above considerations yield the correct picture describing the ground-state physics 
for a closed-system Bose-Hubbard dimer. However, different physics becomes important for an 
\emph{open-system} dimer, in which bosons are not understood as massive 
particles but rather as
excitations that can be created by a coherent drive, as well as disappear from the system by 
energy dissipation. Concrete examples of such a system are coupled nonlinear resonators in 
which photons are the bosonic excitations in question 
\cite{Raftery2014,Fink2017,Fitzpatrick2017}. The HSI mapping we wish to derive 
becomes 
meaningful in this open-system setting, where it relates the nonequilibrium dynamics of the 
positive-$U$ dimer to that of the negative-$U$ dimer.

For our derivation, we assume weak system-bath coupling and validity of the 
Markov approximation, so that we can 
describe the time evolution and steady state 
of the open Bose-Hubbard dimer within the Lindblad master equation formalism 
\cite{breuer2002,Lendi2007}. The reduced density matrices $\orho_{\pm}$ for positive or 
negative $U$ then evolve according to
\be
\label{eq:Master_dimer}
\ds{\orho_{\pm}}{t} = -i \lf[\oH_{\pm}, \orho_{\pm} \rt] + \gar \sum_{n=1}^{2}  
\sD \lf[\oa_n \rt] \orho_{\pm},
\ee
where  $\sD [\oa_n ] \orho \equiv \oa_n \orho \oa_n\dg - \frac{1}{2} \oa_n\dg 
\oa_n \orho - \frac{1}{2} \orho \oa_n\dg \oa_n $ is the dissipator describing 
the non-unitary evolution induced by the system-bath coupling. The jump 
operators $\oa_n$ produce bosonic excitation loss from each site, as is 
appropriate, e.g., to describe intrinsic photon loss in transmission-line 
resonators or optical 
cavities. We remark that \cref{eq:Master_dimer} is widely used to describe the 
open Bose-Hubbard dimer and related models even though, strictly speaking, the 
employed 
jump operators do not obey the requirement that jump operators be operators projecting from 
one eigenstate of the Hamiltonian to another one \cite{breuer2002}. It is worth noting that use of such 
``phenomenological'' dissipators has yielded  quantitative agreement with experimental data 
for  driven-dissipative photonic systems in specific parameter regimes 
\cite{Bishop2009,Raftery2014,Fink2017,Fitzpatrick2017}. A more detailed discussion of this point is beyond the scope of this 
paper.

While both $\oH_{+}$ and $\oH_{-}$ conserve the total number of bosonic excitations, 
dissipation induces relaxation of the dimer towards its vacuum state. 
An external drive can establish a balance between excitation loss and gain. 
For concreteness, we consider a coherent tone driving the first dimer site as described by the 
drive Hamiltonian
$
\oH_{d} (t) =  \ep (\oa_1\dg e^{- i \omd t}  + \text{h.c.}).
$
Here, $\ep$ parametrizes the strength of the drive, and $\omd$ its frequency. In the 
frame co-rotating with the drive,
the effective system Hamiltonian is time independent,
\be
\label{eq:RF_H_dimer}
\oH_{\pm}^\text{eff} = 
\sum_{n=1}^{2} \lf[ \domr \, \oa_n\dg \oa_n  + U_\pm \oa_n\dg \oa_n\dg \oa_n 
\oa_n  \rt] + \ep ( \oa_1\dg + \oa_1 ) + J ( \oa_1\dg \oa_2 + 
\oa_2\dg \oa_1 ),
\ee
where $\domr \equiv \omr - \omd$ denotes the detuning between resonator and 
drive frequency.

\begin{figure}	
	\centering	
	\includegraphics[width=0.85\columnwidth]{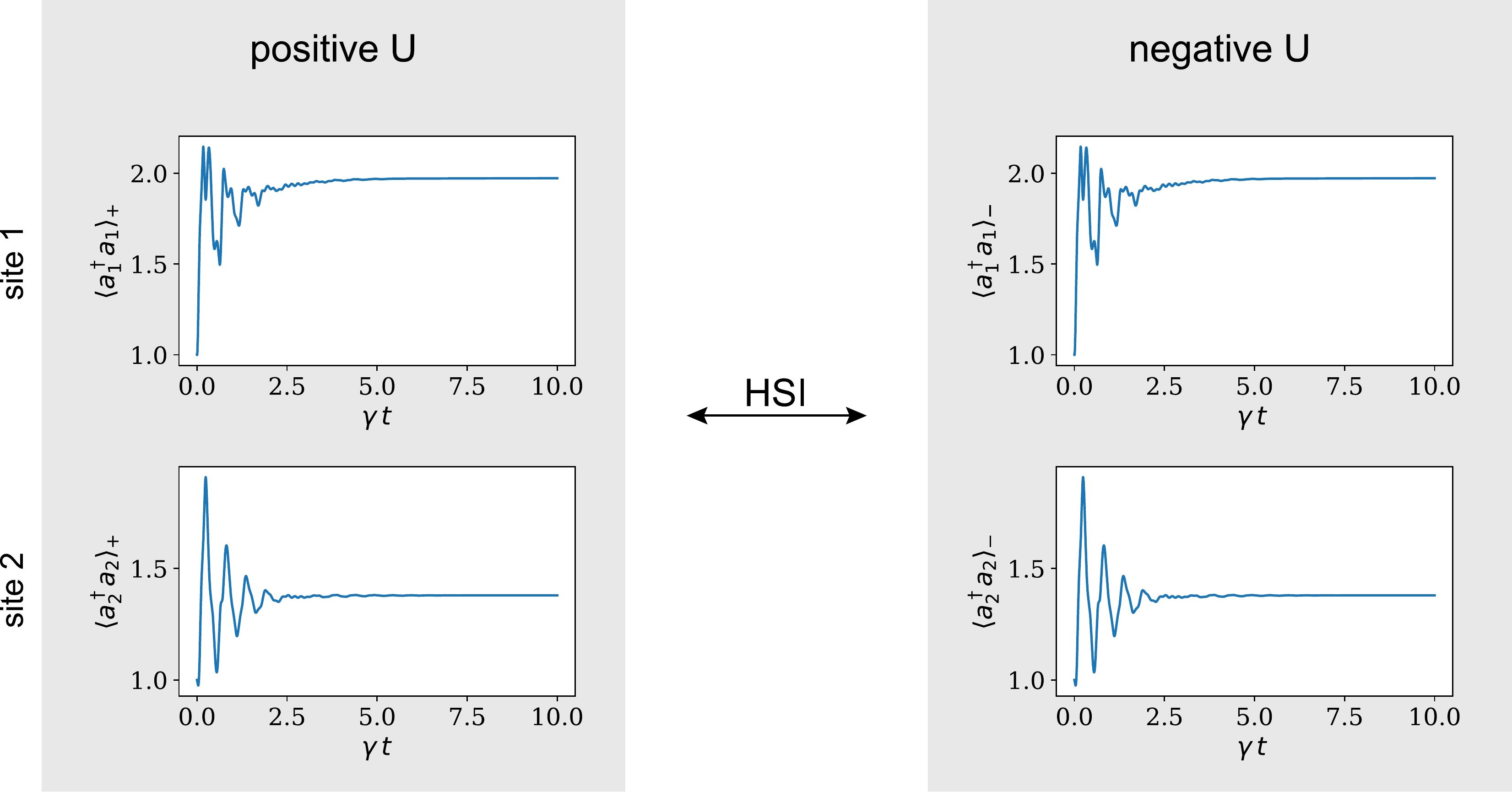}
	\caption{\textbf{Excitation numbers on the two sites of a repulsive and 
	attractive Bose-Hubbard dimer}. The excitation numbers 
		$\savg{\oa_1\dg \oa_1}_{\pm}$ and $\savg{\oa_2\dg \oa_2}_{\pm}$ are 
		observed to follow the same dynamics in the repulsive ($+$) and 
		and attractive case ($-$). For both dimer models, the $t=0$ initial 
		state is the Fock state with one excitation on each site. 
		(Parameters: $\domr/\gar = \pm 1$, $U/\gar = \pm 5$, $\ep/\gar = 15$ and 
		$J/\gar=10$.)}
	\label{fig:dimer_dynamics}
\end{figure}

We demonstrate the HSI mapping at the level of expectation values. Consider for 
instance 
$\savg{\oa_1}$, whose real and imaginary parts yield the two  field quadratures 
$I$ and $Q$ in quantum-optics language. For positive-$U$ and 
negative-$U$ interaction, respectively, the time evolution of 
$\savg{\oa_1}$ is governed by
\begin{align}
\label{eq:a1}
i \ds{ \savg{\oa_1}_\fnpl}{t} &= 
\lf( \domr   - \frac{i \gar}{2} \rt) \savg{\oa_1}_\fnpl
+ 2U_+ \savg{\oa_1\dg \oa_1^2}_\fnpl
+ J  \savg{\oa_2}_\fnpl + \ep,\\
\label{eq:a1minus}
i \ds{ \savg{\oa_1}_\fnpm}{t} &= 
\lf( \domr   - \frac{i \gar}{2} \rt) \savg{\oa_1}_\fnpm
+ 2U_- \savg{\oa_1\dg \oa_1^2}_\fnpm
+ J  \savg{\oa_2}_\fnpm + \ep.
\end{align}
Our claim, to be substantiated in the following, is that the dynamics for
 negative-$U$ interaction can be obtained exactly from the dynamics for 
positive-$U$ interaction. To make this argument, we now consider the 
positive-$U$ system. For 
convenience, we introduce the notation $ 
\savg{\oa_1(\mathfrak{p})}_\fnpl$, where $\mathfrak{p}=(U_+,\,\domr,\, \ep,\, 
J)$ collects all external parameters entering the Hamiltonian 
$\oH_{+}^\text{eff}$ [\cref{eq:RF_H_dimer}]. (Note that we purposely do \emph{not} include the 
dissipation rate $\gamma$  in $\mathfrak{p}$.) Next, we take the complex 
conjugate of \cref{eq:a1} and write it in the form
\begin{align}
\label{eq:a1cc}
i \ds{ \savg{\oa_1(\mathfrak{p})}^*_\fnpl}{t} &= 
\lf(-\domr   - \frac{i \gar}{2} \rt) \savg{\oa_1(\mathfrak{p})}^*_\fnpl
+ 2U_- \savg{\oa_1\dg \oa_1^2(\mathfrak{p})}^*_\fnpl
- J  \savg{\oa_2(\mathfrak{p})}^*_\fnpl - \ep,
\end{align}
where we have used that $-U_+=U_-$.
Comparison with \cref{eq:a1minus} is suggestive of the relation 
\be
\label{eq:a1rel}
\savg{\oa_1(-\mathfrak{p})}_\fnpm =  
\savg{\oa_1(\mathfrak{p})}^*_\fnpl,
\ee
i.e., expectation values for the cases of attractive and repulsive onsite 
interaction are the same up to complex conjugation  and sign adjustments of 
remaining Hamiltonian parameters.
However, a firm proof of this relation 
requires that analogous relations also hold for $\savg{\oa_2}^*_\fnpl$ and 
$\savg{\oa_1\dg 
\oa_1^2}^*_\fnpl$ and thus, due to the ensuing hierarchy of equations of 
motion, for 
all expectation values $\savg{\mathsf{A}^{p,q}_{r,s}}^*_\fnpl=\savg{(\oa_1\dg)^{p} 
(\oa_2\dg)^{q}\,\oa_1^{r}\,\oa_2^{s}}_\fnpl^*$. We show in \cref{app:details_dimer_eom} that 
the relation \eqref{eq:a1rel} indeed carries over to the general case: 
\be\label{eq:Amap}
\savg{\mathsf{A}^{p,q}_{r,s}(-\mathfrak{p})}_\fnpm 
=\savg{\mathsf{A}^{p,q}_{r,s}(\mathfrak{p})}_\fnpl^*.
\ee
In simple words: every expectation value describing the dynamics for negative-$U$ interaction 
can be obtained from a corresponding expectation value for positive-$U$ interaction by the 
following two steps. First, invert the sign of each Hamiltonian parameter, while leaving the 
signs of decoherence rates unchanged. Second, replace expectation values by 
their complex conjugates. The relation \cref{eq:Amap} therefore establishes a one-to-one map 
between positive-$U$ and negative-$U$ interaction through Hamiltonian sign inversion. 
This is 
 summarized by the diagram\\
\be
\label{eq:genuine_mapping_dimer}
	 \includegraphics[width=0.85\columnwidth]{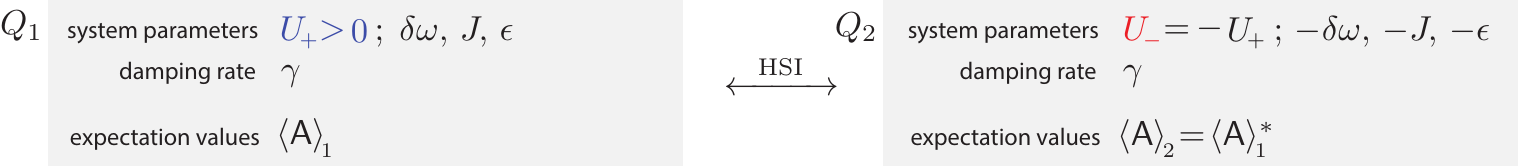}
\ee
where entries in each row specify corresponding Hamiltonian parameters, 
damping parameters, and expectation values.

In order to make the systems $Q_1$ and $Q_2$ with positive and negative $U$  
match even more closely, we may eliminate the 
sign changes in hopping $J$ and drive strength $\epsilon$ with a gauge 
transformation, $\oa_1 \rightarrow 
-\oa_1$. At this point, we find that 
the dynamics of the attractive versus the repulsive driven-damped Bose-Hubbard dimer is 
exactly the same when switching from red-detuned to 
blue-detuned drive frequency, $\domr \to -\domr$. (A similar observation for a driven-damped nonlinear oscillator was made by Dykman in Ref.\ \onlinecite{Dykman2007}.) While the ground-state physics 
of the closed system crucially depends on the sign of the interaction, the 
nonequilibrium dynamics is identical in the discussed sense. We have confirmed this statement  
with multiple numerical simulations. An example of simulation results is depicted in  
\cref{fig:dimer_dynamics}. Here, both positive and negative $U$ dimers are initialized in a 
Fock state with one excitation on each site. The dynamics observed for the  
excitation numbers on the two sites are found to be identical for positive and negative $U$. 
We have confirmed independently that dynamics with different initial states converge to the 
same steady state for positive and negative $U$.

It is interesting to note that the HSI mapping enables one to extend previous results for the 
driven-dissipative Bose-Hubbard dimer to the regime with the opposite sign of interaction. For 
instance for repulsive interaction, it has been predicted that the steady state of the dimer 
can undergo spatial symmetry breaking \cite{Cao2016}. The HSI mapping, then, implies that the 
same symmetry breaking must also be present in the attractive dimer model. Surprisingly, the 
nature of the interaction appears to play only a secondary role in producing the spatial 
symmetry breaking.

\section{Hamiltonian sign inversion mapping}
In order to prove the HSI mapping that links positive-$U$ and negative-$U$ Bose-Hubbard dimers, we invoked the entire 
hierarchy of coupled equations of 
motion for system 
observables. This approach is cumbersome, and leaves one with the question 
whether the HSI mapping relies upon specific properties of the Bose-Hubbard dimer, which would limit its scope to this one particular model. We will demonstrate that this is not 
the case and show that, rather, the HSI mapping generalizes to arbitrary Markovian open 
quantum systems with time-reversal invariant system Hamiltonians $\oH$. 
(Note that while $\oH$ may be time-reversal invariant, the coupling of the 
system to its environment will naturally break overall time-reversal symmetry.) 
The HSI mapping establishes a one-to-one correspondence 
between the dynamics of an open quantum system $Q_1$  with system Hamiltonian 
$\oH$ and the dynamics of a partner system $Q_2$ with system Hamiltonian 
$-\oH$. We will base our discussion on the 
Lindblad master equation, and show that the HSI mapping can be formulated in a 
straightfoward way that entirely bypasses cumbersome considerations of the 
hierarchy of equations of motion.

The dynamics of of the open system $Q_1$ is governed by the Lindblad master 
equation 
\cite{breuer2002,Lendi2007},
\be
\label{eq:master_eq}
\ds{}{t} \orho(t) =  -i \lf[ \oH,  \orho(t) \rt] + \sum_{j} \ga_j 
\sD[\occ_j] \orho(t),
\ee
which describes the time evolution $t\mapsto\orho(t)$ of the reduced density 
matrix of $Q_1$. The dissipation and dephasing processes from coupling 
to the environment are encoded by rates $\ga_j$ and corresponding 
jump operators $\occ_j$. In the absence of coupling to the environment, the system is assumed to be
time-reversal symmetric. As usual, we formalize this symmetry by utilizing the 
antiunitary time-reversal operator $\oT$ \cite{Wigner1959,Bargmann1964},  which 
must be constructed for each concrete system of interest so that relevant 
observables obey the appropriate transformation laws, such as 
 $\oT \mathsf{x} \oT\dg=\mathsf{x}$ and 
$\oT \mathsf{p} \oT\dg=-\mathsf{p}$ for generalized position  and conjugate 
momentum  operators $\mathsf{x}$ and $\mathsf{p}$. Time-reversal symmetry of 
the isolated 
system then amounts to the identity  $\oT\oH\oT\dg = \oH$.

We construct the general HSI mapping by considering the $\oT$-transform of the 
density 
matrix,
\be
\orho_T = \oT \orho \oT\dg.
\ee
We stress that the evolution $t\mapsto\orho_T(t)$ does \emph{not} 
correspond to backward-in-time evolution of $t\mapsto\orho(t)$. We obtain the 
equation of motion for  $\orho_T(t)$
by sandwiching \cref{eq:master_eq} with $\oT$ and $\oT\dg$, exploiting that the
time-reversal operator obeys $\oT\dg\oT=\openone$, and invoking time-reversal 
symmetry of the system Hamiltonian. This yields the equation
\be
\label{eq:reversed_master_eq}
\ds{}{t} \orho_T(t) = -i \lf[ -\oH,  \orho_T(t) \rt] + \sum_{j} \ga_j \sD[ \oT 
\occ_j \oT\dg ] \orho_T(t).
\ee
which we recognize as having the proper form of a Lindblad master equation. 
Hence, we may 
interpret $\orho_T$ as the density matrix of an open quantum system $Q_2$. 
Comparing this \cref{eq:reversed_master_eq} with the original master equation  
\eqref{eq:master_eq} for $\orho$, we see that 
$Q_2$ has a Hamiltonian with inverted 
sign, as well as jump operators $\oT \occ_j \oT\dg$. 

We can now relate expectation 
values $\savg{\oO}_2 = \Tr (\oO\orho_T)$ for system $Q_2$ back 
to expectation values for $Q_1$.
To do so, write $\savg{\oO}_2 = \Tr (\oT\oT\dg\oO\oT\orho\oT\dg)$, but note 
that the 
cyclic property of the trace does \emph{not} hold for 
anti-linear 
operators such as $\oT$. Instead, we simplify the expression further by 
considering an 
orthonormal Hilbert space basis of time-reversal invariant states 
$\{\ket{n}\}$. In this 
basis, the action of the 
time-reversal operator reduces to complex 
conjugation, such that $\oT \sum_n\alpha_n\ket{n}  = 
\mathcal{K}\sum_n\alpha_n\ket{n}=\sum_n\alpha_n^*\ket{n}$. With this, we find
\be
\savg{\oO}_2 = 
\Tr(\oT\bar{\oA}\orho\oT\dg) = \sum_n 
\bra{n}\mathcal{K}\bar{\oA}\orho\mathcal{K}\ket{n} = 
 \sum_{m,n} 
\bra{n}\mathcal{K} \langle m | \bar{\oA}\orho | n\rangle\ket{m} = \Tr( 
\oT\dg\oA\oT\orho)^*,
\ee
where we have temporarily used the shorthand $\bar{\oA}=\oT\dg \oA \oT$.
As a result, the correspondence between expectation values in system $Q_1$ 
and $Q_2$  takes the form 
\be
\savg{\oO}_2  =  \Tr (\oO\orho_T) =  \Tr (\oT\dg\oO\oT\orho)^* =  
\savg{\oT\dg\oA\oT}_1^*.
\ee
We can summarize the general HSI mapping with the diagram
\be
\label{eq:TRM_general_1}
	\includegraphics[width=0.8\columnwidth]{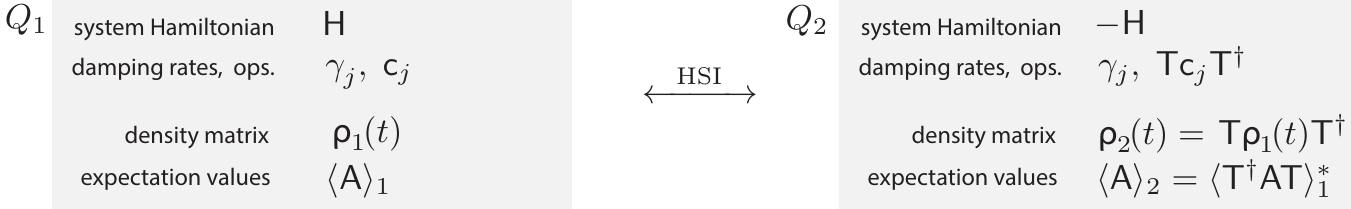}
\ee

It is easy to verify that the mapping \eqref{eq:genuine_mapping_dimer} for the 
driven-dissipative Bose-Hubbard dimer is a 
special case of \eqref{eq:TRM_general_1}. To see this note that: the inversion 
of the Hamiltonian sign 
produces the sign changes of system parameters as recorded in 
\cref{eq:genuine_mapping_dimer}; the jump operators considered in the dimer 
model are time-reversal invariant, $\oT\oa_j\oT\dg=\oa_j$ \footnote{The 
annihilation operator is defined by $\oa = \tfrac{1}{\sqrt{2}} 
(\frac{\mathsf{x}}{x_0} + i \frac{\mathsf{p}}{p_0})$ where $\mathsf{x}$ is a 
generalized position and $\mathsf{p}$ a generalized momentum operator; $x_0$, 
$p_0$ are constants. From $\oT \mathsf{x} \oT\dg=\mathsf{x}$,  $\oT \mathsf{p} 
\oT\dg=-\mathsf{p}$, and anti-linearity of $\oT$ follows the time-reversal 
invariance of $\oa_j$.}, and so are the expectation values of the observables 
$\oA^{p,q}_{r,s}$. We note that the general HSI mapping 
immediately extends the correspondence between positive and negative $U$ 
dimers to Bose-Hubbard lattices of arbitrary size and lattice geometry.

The HSI mapping is mathematically rigorous, but one must check that the partner 
system $Q_2$ is indeed a physically meaningful quantum system. Two aspects are 
crucial here. First, in infinite-dimensional Hilbert spaces, sign-inversion of 
the Hamiltonian leads to energy spectra not bounded from below. As shown in the 
dimer example, this is unproblematic if the Hamiltonian is an effective 
Hamiltonian in a rotating frame, whose eigenvalues only carry the meaning of 
quasienergies. Second, the HSI mapping may modify the jump operators entering 
the master equation.

A wide range of driven open quantum systems is amenable to the HSI mapping, including circuit-QED and ultracold-atoms systems which are of interest in studies of phase transitions \cite{Lee2012,Carmichael2015, Casteels2017,Fink2017,Fitzpatrick2017} and quantum state preparation \cite{Diehl2008,Reiter2013,Aron2014,Kimchi-Schwartz2016}. 
For open quantum systems with finite-dimensional Hilbert space, $-\oH$ is 
always physical. This class of system covers a number of quantum systems 
currently being researched, e.g.\ open spin lattices 
\cite{Prosen2010b,Lee2011,Lee2012,Cai2013,Schwager2013} for which we will 
present one example in the following section. Here, again, the HSI mapping will 
link two physically different systems and establish a useful one-to-one 
correspondence between their nonequilibrium dynamics.

\section{Driven-dissipative spin lattice}
The HSI mapping can easily be applied to driven-dissipative (pseudo-)spin lattices which can be realized, for example, by 
ultracold atoms \cite{Schwager2013} and circuit-QED devices \cite{Nissen2012,Viehmann2013}. 
We illustrate such an application next, considering an Ising system with one spin per lattice site, each with (Zeeman-)energy splitting $\omq$. Each spin 
is driven by a coherent tone with drive strength $\ep_j$ and frequency $\omd$, 
and is $\sigma^z$-coupled to its nearest neighbors with a coupling strength $J$. The 
effects of the environment are modeled by spin relaxation with a rate 
$\gaq$. In the frame co-rotating with the drive, the system dynamics is 
governed by the master equation
\be
\ds{\orho}{t} = 
-i [\oH, \orho]
+\gaq \sum_{j=1}^{N} \sD\lf[ \om_j \rt] \orho,
\ee
with system Hamiltonian
\be
\oH_{\pm}=
\sum_{j=1}^{N} 
\lf[ \domq   \op_j \om_j
+\ep_j \lf( \om_j + \op_j  \rt)
\rt]
\pm J \sum_{\lf< j,k \rt>}  \oz_j \oz_k,
\ee
where $\domq  \equiv \omq - \omd$ is the detuning.

The Ising-coupling strength $J$ can be designed to be positive (anti-ferromagnetic coupling) or negative (ferromagnetic coupling), depending on the particular physical realization \cite{Viehmann2013}.
We consider the special case where the underlying lattice is not bipartite, such as a triangular or Kagome lattice. In this case, ferromagnetic and antiferromagnetic coupling are well known to lead to very different equilibrium physics: while for $\omq = 0$ the negative-$J$ ground state is a simple ferromagnet, the positive-$J$ case faces geometrical frustration of the antiferromagnetic coupling  -- a situation of great interest in many-body physics, e.g.\ in the study 
of spin glasses \cite{Ramirez1994}.
Despite the dramatically different ground-state physics of the geometrically 
frustrated and non-frustrated  lattices, one finds that the HSI mapping 
\eqref{eq:TRM_general_1} establishes a one-to-one correspondence for the 
out-of-equilibrium dynamics under driving and damping. Similar to $\oa_j$ in
the harmonic oscillator case, one verifies that the spin lowering operator $\om_j$ for pseudospins is invariant under time reversal. Hence, it is straightforward to apply the HSI 
mapping and obtain
\be
\includegraphics[width=0.8\columnwidth]{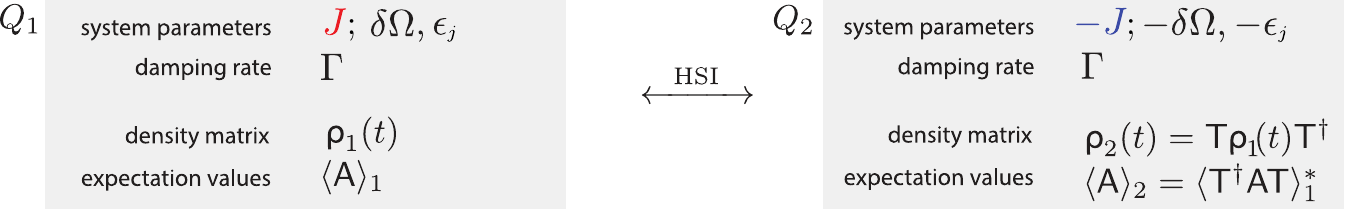}
\ee
where system $Q_1$ is the frustrated spin lattice, $Q_2$ the non-frustrated 
lattice.
We can make the Hamiltonian of $Q_2$ match the one in $Q_1$ even more closely 
by eliminating the sign 
change in the drive strength $\ep_j$ with a gauge transformation, $\ox_j 
\rightarrow -\ox_j$ and $\oy_j \rightarrow - \oy_j$ for all sites $j$.
As a result of this, we find a one-to-one correspondence between the frustrated 
and non-frustrated spin dynamics and steady state in a driven-dissipative Ising
lattice. (The only parameter to be adjusted is the drive-frequency detuning 
$\delta\Omega$.)

\begin{figure}	
	\centering	
	\includegraphics[width=0.9\columnwidth]{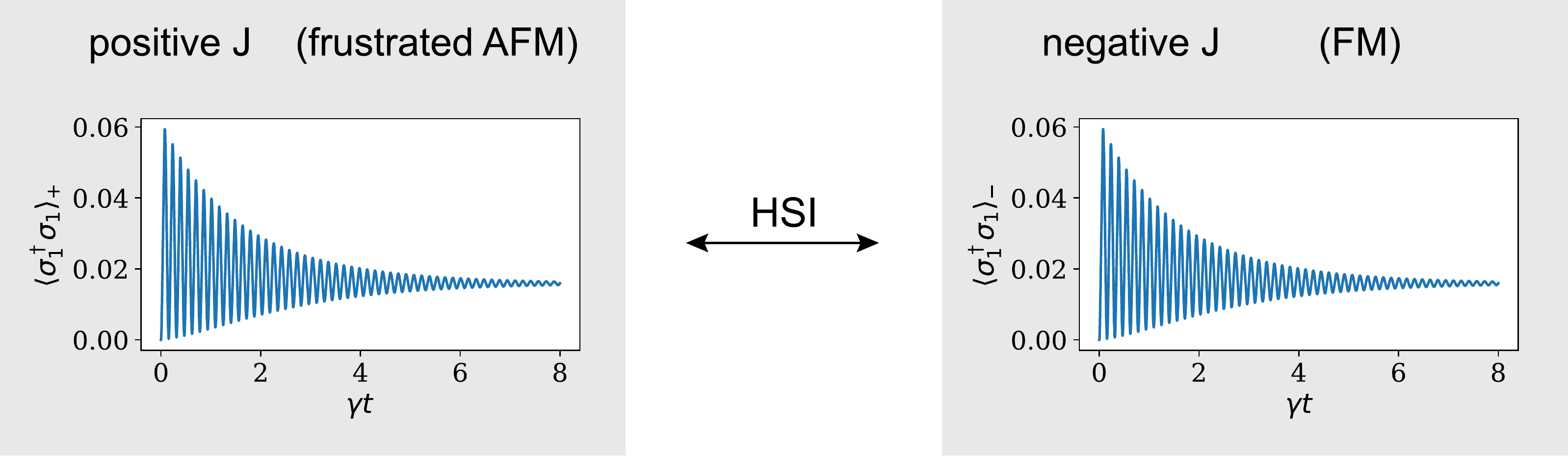}
	\caption{\textbf{Dynamics of spin excitation numbers for a triangular plaquette of Ising-coupled spins subject to driving and dissipation, with antiferromagnetic and ferromagnetic coupling}. 
	The excitation number $\savg{\op_1 \om_1}_{\pm}$ is observed to follow 
	the same dynamics for the frustrated ($+$)  and non-frustrated ($-$) 
	case. (Parameters: $\domq/\gar = \pm 1$, $\ep/\gar = 5$ and $J/\gar= \pm 
	5$; initial state: all spins in ground state.)}
	\label{fig:spin_lattice_evolution}
\end{figure}

For numerical confirmation of this result we have simulated the dynamics for a 
triangular plaquette of three spins, see \cref{fig:spin_lattice_evolution}.  
(In this 
simulation, only one of the three sites is driven and the spin 
excitation 
number on that particular site is monitored.) We find the expected, but 
non-intuitive, result that the nonequilibrium dynamics are indeed identical in 
the frustrated and the non-frustrated case. 

\section{Conclusion}
We have proven and illustrated the use of a mapping that establishes a 
one-to-one correspondence between nonequilibrium dynamics of one Markovian open 
quantum  system and a second such system whose Hamiltonian carries the opposite 
sign. This mapping relies on the time-reversal invariant of the system Hamiltonian, and makes the remarkable prediction that nonequilibrium 
dynamics of different systems can be essentially identical despite the fact 
that their equilibrium physics is extremely different. We demonstrated this 
Hamiltonian sign inversion mapping for two concrete examples: the 
driven-dissipative Bose-Hubbard dimer for attractive vs.\ repulsive onsite 
interaction, and a driven-dissipative Ising spin lattice model with and without 
geometrical frustration.

The HSI mapping is widely applicable to many interesting driven-dissipative 
quantum models realizable by ultracold atoms and circuit-QED architecture, and 
allows 
important conclusions. For instance, from the HSI 
mapping we can immediately infer that the symmetry-breaking state predicted for 
the driven-dissipative 
\emph{repulsive} Bose-Hubbard dimer \cite{Cao2016} must also occur for the case 
of an 
\emph{attractive} dimer. Establishing such mappings will not only facilitate a 
better 
understanding of the relation between broken symmetry and the nature of  
nonlinearities, but also motivate new experiments probing nonequilibrium 
many-body phenomena. We believe that the HSI mapping will serve as a valuable 
tool in 
the study of nonequilibrium dynamics, steady-state properties, and dissipative 
phase transitions in open quantum systems.

\begin{acknowledgments}
We acknowledge valuable discussions with Peter Groszkowski, Mattias Fitzpatrick, Neereja Sundaresan, A.\ A.\ Houck and M.\ I.\ Dykman. This research was supported in part by the NSF under Grant No.\ PHY-1055993 (A.C.Y.L.\ and J.K.).
\end{acknowledgments}

\appendix
\section{Equations of motion of the driven-dissipative Bose-Hubbard dimer model
\label{app:details_dimer_eom}}
In this appendix, we demonstrate that expectation values of the general operators 
$\mathsf{A}^{p,q}_{r,s}=(\oa_1\dg)^{p} (\oa_2\dg)^{q}\,\oa_1^{r}\,\oa_2^{s}$
indeed obey the relation
\be\label{eq:Arefapp}
\savg{\mathsf{A}^{p,q}_{r,s}(-\mathfrak{p})}_\fnpm 
=\savg{\mathsf{A}^{p,q}_{r,s}(\mathfrak{p})}_\fnpl^*,
\ee
see \cref{eq:Amap} in the main text.
We abbreviate the positive-$U$ expectation value by $A^{p,q}_{r,s} = 
\savg{\mathsf{A}^{p,q}_{r,s}(\mathfrak{p})}_\fnpl$ and deduce the corresponding equation of 
motion from the master equation \eqref{eq:Master_dimer}. We find:
\begin{align}
&i \ds{A^{p,q}_{r,s} }{t} = \nn\\
&\quad\lf[ \lf(\domr - U\rt) \lf(r + s - p - q \rt) + U \lf( r^2 + s^2 - p^2 - 
q^2 \rt)  - \frac{i \gar}{2} \lf( p+q+r+s \rt)\rt]A^{p,q}_{r,s}
 + 2U \lf[ (r-p)A^{p+1,q}_{r+1,s} + (s-q)A^{p,q+1}_{r,s+1} \rt]
\nn\\
&\quad + J \lf[ r A^{p,q}_{r-1,s+1} + s A^{p,q}_{r+1,s-1} - p A^{p-1,q+1}_{r,s} 
-q 
A^{p+1,q-1}_{r,s} \rt]
 + \ep \lf[ r A^{p,q}_{r-1,s} - p  A^{p-1,q}_{r,s} \rt].
\end{align}
By complex conjugation of this equation, we obtain
\begin{align}
&i \ds{A^{p,q*}_{r,s}}{t} = \nn\\
&\, \lf[ \lf(- \domr + U\rt) \lf(r + s - p - q \rt) - U \lf( r^2 + s^2 - p^2 - 
q^2 \rt)  - \frac{i \gar}{2} \lf( p+q+r+s \rt)\rt] A^{p,q*}_{r,s}
 - 2U \lf[ (r-p) A^{p+1,q*}_{r+1,s} + (s-q) A^{p,q+1*}_{r,s+1}\rt]
\nn\\
&\, - J \lf[ r A^{p,q*}_{r-1,s+1} + s A^{p,q*}_{r+1,s-1} - p A^{p-1,q+1*}_{r,s} 
- q 
A^{p+1,q-1*}_{r,s} \rt]
 - \ep \lf[ r A^{p,q*}_{r-1,s} - p A^{p-1,q*}_{r,s} \rt].
\end{align}
Comparison of the latter equation with the equation of motion for the corresponding 
negative-$U$ expectation value confirms the proposed relation 
\eqref{eq:Arefapp}.

\bibliographystyle{./apsrev4-1-prx}
\bibliography{cit}
\end{document}

%% file: header_command.tex
\newcommand{\be}{\begin{equation}}
\newcommand{\ee}{\end{equation}}
\newcommand{\nn}{\nonumber}
\newcommand{\lf}{\left}
\newcommand{\rt}{\right}

\newcommand{\Tr}{\mathrm{Tr}}

\newcommand{\dg}{^{\da}}

\newcommand{\da}{\dagger}

\newcommand{\ep}{\epsilon}
\newcommand{\ga}{\gamma}

\newcommand{\omd}{\omega_{d}}
\newcommand{\domr}{\delta \omega}
\newcommand{\domq}{\delta \Omega}
\newcommand{\omr}{\omega}
\newcommand{\omq}{\Omega}
\newcommand{\gar}{\gamma}
\newcommand{\gaq}{\Gamma}

\newcommand{\ds}[2]{\frac{\mathrm{d} {#1} }{\mathrm{d} {#2}} }

\newcommand{\bra}[1]{\lf< #1 \rt|}
\newcommand{\ket}[1]{\lf| #1 \rt>}

\newcommand{\savg}[1]{\langle #1 \rangle}

\newcommand{\op}{\osigma^{+}}
\newcommand{\om}{\osigma^{-}}
\newcommand{\ox}{\osigma^{x}}
\newcommand{\oy}{\osigma^{y}}
\newcommand{\oz}{\osigma^{z}}

\newcommand{\oa}{\mathsf{a}}
\newcommand{\oA}{\mathsf{A}}

\newcommand{\occ}{\mathsf{c}}

\newcommand{\oH}{\mathsf{H}}

\newcommand{\oO}{\oA}

\newcommand{\orho}{\text{\textrho}}
\newcommand{\osigma}{\text{\textsigma}}

\newcommand{\oT}{\mathsf{T}}

\newcommand{\sD}{\mathbb{D}}

